\documentclass[runningheads]{llncs}
\pdfoutput=1
\usepackage{graphicx}
\usepackage{microtype}
\usepackage{xspace}
\usepackage{xcolor}
\usepackage{amsmath}
\usepackage[capitalise]{cleveref}
\usepackage{booktabs}
\usepackage{array}
\usepackage{subfig}
\usepackage{placeins}
\usepackage{paralist}
\usepackage[%
  group-minimum-digits=4,%
  list-final-separator={, and },%
  add-integer-zero=false,%
  free-standing-units,%
  round-mode=figures,%
  round-precision=3,%
  binary-units,%
  detect-weight=true,%
  detect-inline-weight=math,%
]{siunitx}

\usepackage{listings}
\usepackage{cite}

\newcommand\definetool[2]{\newcommand{#1}{{\textsc{#2}}\xspace}}
\definetool{\scratch}{Scratch}
\definetool{\whisker}{Whisker}
\definetool{\servant}{Servant}

\definecolor{commentgray}{rgb}{0.31,0.31,0.31}
\definecolor{codegray}{rgb}{0.5,0.5,0.5}
\definecolor{codepurple}{rgb}{0.58,0,0.82}
\definecolor{backcolour}{rgb}{0.94,0.95,0.96}
\definecolor{darkgreen}{rgb}{0.11,0.520.21}

\lstdefinelanguage{JavaScript}{
  morekeywords=[1]{break, continue, delete, else, for, function, if, in,
    new, return, this, typeof, var, void, while, with},
  morekeywords=[2]{false, null, true, boolean, number, undefined,
    Array, Boolean, Date, Math, Number, String, Object},
  morekeywords=[3]{eval, parseInt, parseFloat, escape, unescape},
  sensitive,
  morecomment=[s]{/*}{*/},
  morecomment=[l]//,
  morecomment=[s]{/**}{*/}, % JavaDoc style comments
  morestring=[b]',
  morestring=[b]"
}[keywords, comments, strings]
\lstalias[]{ES6}[ECMAScript2015]{JavaScript}
\lstdefinelanguage[ECMAScript2015]{JavaScript}[]{JavaScript}{
  morekeywords=[1]{await, async, case, catch, class, const, default, do,
    enum, export, extends, finally, from, implements, import, instanceof,
    let, static, super, switch, throw, try},
  morestring=[b]` % Interpolation strings.
}

\lstdefinestyle{code}{
  backgroundcolor=\color{backcolour},
  commentstyle=\color{commentgray},
  keywordstyle=\color{violet},
  numberstyle=\tiny\color{codegray},
  stringstyle=\color{darkgreen},
  basicstyle=\linespread{1.1}\ttfamily,
  breakatwhitespace=false,
  breaklines=true,
  captionpos=b,
  keepspaces=true,
  numbers=left,
  numbersep=5pt,
  showspaces=false,
  showstringspaces=false,
  showtabs=false,
  tabsize=2,
  frame=tlbr,
  framerule=0pt,
}
\newcommand{\event}[1]{{\sffamily #1}}

\begin{document}
\title{Search-based Testing for Scratch Programs}
%
%\titlerunning{Abbreviated paper title}
% If the paper title is too long for the running head, you can set
% an abbreviated paper title here
%
\author{Adina Deiner \and
Christoph Fr\"{a}drich \and \\
Gordon Fraser \and
Sophia Geserer \and
Niklas Zantner\thanks{Authors listed in alphabetical order.}}
\authorrunning{Deiner et al.}

\institute{University of Passau, Innstr. 33, 94032 Passau, Germany}

\maketitle              % typeset the header of the contribution
\begin{abstract}
	%
	% Context
	%
	Block-based programming languages enable young learners to quickly implement fun programs and games. The \scratch programming environment is particularly successful at this, with more than 50 million registered users at the time of this writing.
	%
	% Problem
	%
	Although \scratch simplifies creating syntactically correct programs, learners and educators nevertheless frequently require feedback and support. Dynamic program analysis could enable automation of this support, but the test suites necessary for dynamic analysis do not usually exist for \scratch programs.
	%
	% Insight
	%
	It is, however, possible to cast test generation for \scratch as a search problem.
	%
	% Contribution
	%
	In this paper, we introduce an approach for automatically generating test suites for \scratch programs using grammatical evolution. The use of grammatical evolution clearly separates the search encoding from framework-specific implementation details, and allows us to use advanced test acceleration techniques.
	%
	% Results
	%
	We implemented our approach as an extension of the \whisker test framework. Evaluation on sample \scratch programs demonstrates the potential of the approach.
\end{abstract}
%
%
%

%%%%%%%%%%%%%%%%%%%%%%%%%%%%%%%%%%%%%%%%%%%%%%%%%%%%%%%%%%%%%%%%%%%%%%%%%%%%%%%
\section{Introduction}
\label{sec:introduction}

Visual, block-based programming languages are a popular means to introduce young learners to programming. Programs are created by visually arranging   high-level program statements, thus allowing learners to quickly and easily create engaging programs and fun games. There are many different programming environments built on this idea, and the \scratch~\cite{maloney2010scratch} programming environment is one of the most popular of these, with more than 50 million registered users at the time of this writing\footnote{\url{https://scratch.mit.edu/statistics/}, last accessed 9.6.2020}. While the visual representation ensures that statements can only be assembled in syntactically valid ways, achieving desired functionality can nevertheless be challenging: Testing, debugging, and fixing programs can challenge learners, as well as educators who may aim to support or assess them.
In regular programming, this support is often provided by dynamic analysis: Given a test suite, one can check the runtime properties of a program, determine whether functionality is satisfied, and locate possible causes of failures. However, test suites for \scratch programs do not typically exist.

The \whisker testing framework~\cite{TestingScratch} has been introduced as a means to automate testing of \scratch programs. \whisker-tests interact with \scratch programs through the user interface, for example by providing key-presses and mouse-clicks as inputs. In \whisker, these inputs have to be scripted by a tester, or can be generated randomly. While \scratch programs are often trivially simple, this is not always the case, thus challenging the test generator. We therefore cast the problem of generating \whisker tests as search problem.

We use many-objective optimisation to evolve sets of tests that cover as many as possible program statements. We use grammatical evolution, where search is applied to an integer-list representation using traditional search operators, and the integers are decoded to UI events using a dynamically generated input grammar for the \scratch program under test. Fitness evaluation requires test execution, which is challenging because (1) \scratch programs are UI-centric and often encode timed behaviour, thus making test execution slow, and because (2) \scratch programs are interpreted by the \scratch virtual machine (VM) based on a custom internal representation, thus making program instrumentation difficult. Our implementation overcomes these challenges by integrating an accelerated, headless execution framework to speed up test generation.

In detail, the contributions of this paper are as follows:
\begin{itemize}
	\item We cast \scratch testing as a many-objective search problem using grammatical evolution (\cref{sec:evolution}) and many-objective search (\cref{sec:algorithm}).
	\item We define coverage-based fitness functions (\cref{sec:fitness}) and provide efficient means to evaluate fitness (\cref{sec:execution}).
	\item We illustrate the problem domain and potential of the approach using example \scratch programs (\cref{sec:evaluation}).
\end{itemize}

%\todo{Outline main findings}
%\todo{Mention open source}

%%%%%%%%%%%%%%%%%%%%%%%%%%%%%%%%%%%%%%%%%%%%%%%%%%%%%%%%%%%%%%%%%%%%%%%%%%%%%%%
\section{Background}
\label{sec:background}

%------------------------------------------------------------------------------
\subsection{\scratch Programs}

\newcommand{\ctrllocations}{\ensuremath{L}\xspace}%
\newcommand{\datalocations}{\ensuremath{X}\xspace}%
\newcommand{\dataloc}{\ensuremath{x}\xspace}%
\newcommand{\controlflows}{\ensuremath{G}\xspace}%
\newcommand{\scripts}{\ensuremath{S}\xspace}%
\newcommand{\locations}{\ctrllocations}%
\newcommand{\ops}{\ensuremath{Ops}\xspace}%
\newcommand{\location}{\ensuremath{l}\xspace}%
\newcommand{\process}{\ensuremath{p}\xspace}%
\newcommand{\processes}{\ensuremath{P}\xspace}%
\newcommand{\concrete}{\ensuremath{c}\xspace}%
\newcommand{\concretes}{\ensuremath{C}\xspace}%
\newcommand{\program}{\ensuremath{{\mathrm{App}}}\xspace}%
\newcommand{\step}{\textsf{step}\xspace}
\newcommand{\executiontrace}{\ensuremath{\bar \concrete}\xspace}%

A \scratch program consists of the \emph{stage}, which represents the application window and background image, and a collection of \emph{sprites} that are rendered as different images on top of the stage. The stage and each of the sprites contain a number of \emph{scripts} $\scripts$ that define the program logic. Scripts are created by visually arranging \emph{blocks} that correspond to syntactical elements of the language, such as control-flow structures or expressions.
A \emph{script}~$s = (\ctrllocations, \datalocations, \controlflows, l_0) \in
\scripts$ is a tuple that represents a control-flow automaton, with the set of
\emph{control locations}~$\ctrllocations$, the set of \emph{data
locations}~$\datalocations$, the \emph{control transition
relation}~$\controlflows \subseteq \locations \times \ops \times \locations$ with possible program operations~$\ops$, and the \emph{initial control location}~$\location_0$.
Executing a \scratch program results in the creation of a collection of
concurrent \emph{processes}~$\processes$, each process~$\process \in \processes$ being an instance of a script $s$.
%. Each process~$\process \in \processes$ is the instance of a script~$s = (\ctrllocations, \datalocations, \controlflows, l_0) \in \scripts$. 
%
A \emph{concrete state}~$\concrete = \langle \process_1, \ldots, \process_n \rangle \in \concretes$
of a \scratch program is modelled as a list of concrete process states, which map a concrete value to each data location~$\dataloc \in \datalocations_p$.
%
%An \emph{execution trace}~$\executiontrace_i = \langle \concrete_1, \ldots,
%\concrete_n \rangle$ starts in an \emph{initial concrete state}~$\concrete_1$,
%for which all processes are on the initial control location~$\location_0$ of
%the corresponding scripts. Execution traces describe interleavings of processes that are executed in parallel.
%
\scratch programs are controlled by the user, using mouse, keyboard, or microphone. That is, a program can react to mouse movement, mouse button presses, keyboard key presses, sound levels, or entering answers to ``ask''-blocks. Typically, the first statement of a script $s$ is an event handler block (\emph{hat block}) that links the execution of the script to the occurrence of an event (user events or internally triggered events, such as broadcasts or clone events).

%------------------------------------------------------------------------------
\subsection{The \whisker Test Framework}

\whisker~\cite{TestingScratch} is an automated testing framework for \scratch
programs. A \whisker test consists of a \emph{test harness}, which takes the
role of stimulating the \scratch program under test with inputs, and a set of
\emph{\scratch observers}, which encode properties that should be checked on
the program under test. To execute tests, the \whisker virtual machine wraps
the \scratch virtual machine and its \step-function. Before each execution
step, the test harness is used to produce inputs that are sent as
messages/events to the \scratch program under test, and after each execution of
\step the \scratch observers check whether the resulting state satisfies the
given properties. Each invocation of \step executes the processes concurrently,
and it is possible that several control-flow transitions are taken in one step.
\whisker supports \emph{static test harnesses}, where the system is stimulated with inputs encoded in JavaScript, or \emph{dynamic test harnesses}, where the system is stimulated with randomly determined sequences of inputs. Although random inputs are often sufficient to fully cover simple programs, previous work has shown~\cite{TestingScratch} that programs are not always fully covered. Therefore, the aim of this paper is to use metaheuristic search to automatically generate static test harnesses, i.e., test suites that reach all statements of a program under test.

%%%%%%%%%%%%%%%%%%%%%%%%%%%%%%%%%%%%%%%%%%%%%%%%%%%%%%%%%%%%%%%%%%%%%%%%%%%%%%%
\section{Search-based Testing for Scratch}
\label{sec:sbst}

%------------------------------------------------------------------------------
\subsection{Encoding \scratch Tests Using Grammatical Evolution}
\label{sec:evolution}

Grammatical evolution~\cite{o2001grammatical} (GE) describes a form of Genetic Algorithms (GAs) where the mapping from genotype to phenotype is performed using a problem-specific grammar $G = \langle T, N, P, n_s\rangle$: $T$ is a set of terminals, which are the items that will appear in the resulting phenotype; $N$ are non-terminals, which are intermediate elements associated with the production rules $P: N \rightarrow (N \cup T)*$. The element $n_s \in N$ is the start symbol, which is used at the beginning of the mapping process.

The genotype is typically represented simply as a list of integers (\emph{codons}).
The mapping of a list of codons to the phenotype creates a derivation of the grammar as follows: Beginning with the first production of starting symbol $n_s$ of the grammar, for each non-terminal $x$ on the right hand side of the production we choose the $r$th production rule out of all $n$ rules available for $x$. Given a codon $c$ and $n$ productions for non-terminal $x$, the number $r$ of the production rule to choose is determined as follows:
\begin{equation*}
  r  = c \; mod \; n
\end{equation*}
Each time a production rule is selected, the decoding moves on to the next codon of the genotype. If the end of the genotype has been reached and there are non-terminals left, then usually the selection of codons starts over at the beginning of the genotype.

In order to instantiate GE, we need to define a grammar that represents these tests. A test case is a sequence of user inputs (UI events). Thus, the starting production for a test case of length $n$ is given by the following:
{
  $$
      \begin{aligned}
         testcase ::= \; & input_1 \; input_2 \; \ldots \; input_n
      \end{aligned}
  $$
}
User inputs can be events sent via mouse, keyboard, or microphone. Therefore, terminals in the test grammar will denote concrete mouse, keyboard, or sound events, which may be parameterised. The following grammar thus defines possible inputs:
{
  $$
      \begin{aligned}
         input ::= \; & \textsf{determineEvents}(S, C)
      \end{aligned}
  $$
}
The function $\textsf{determineEvents}(S, C)$ returns a list of all events that the \scratch program consisting of scripts $S$ with a given concrete state $C$ supports. In particular, we support the following events:
\begin{itemize}
	\item \event{KeyPress}: One \event{KeyPress} event is created for each key for which an event handler exists in $S$.
	\item \event{KeyDown}: One \event{KeyDown} event is created for each key for which a key-sensing block exists in $S$.
	\item \event{ClickSprite}: One \event{ClickSprite} event is created for each sprite that contains a click-event handler. Furthermore, for each such sprite we create an additional \event{ClickSprite} event for each clone that exists of the sprite.
	\item \event{ClickStage}: If the stage has a click event handler, then this event is created.
	\item \event{TypeText}: If $S$ uses the $answer$ block and sensing is active at state $C$, then we create one \event{TypeText} event for each concrete string contained in the program. % Random?
	\item \event{MouseDown}: If $S$ contains a sensing mouse button block, this event is added, which toggles the state of the mouse button.
	\item \event{MouseMove}: If $S$ contains a sensing mouse position block, this event is added. The $x$ and $y$ location of the move are determined by the two following codons in the genotype (i.e., \event{MouseMove} implicitly contains a production with two further non-terminals for the coordinates).
	\item \event{Sound}: If $S$ contains event handlers that check the loudness, then for each handler one such event is created, parameterised with the volume checked in the event handler.
	\item \event{Wait}: We always create a Wait event with the default step duration. In addition, one \event{Wait} event is created for each distinct delay value in $S$ (e.g., parameters of wait, say, think, or glide).
\end{itemize}
\scratch programs usually contain main scripts triggered by the \emph{Greenflag} event. We trigger the \emph{Greenflag} event at the beginning of each test, and therefore do not include it in the grammar.
Although the grammar for tests is quite simple, GE offers large implementation benefits as it cleanly separates search operators from the phenotype. For example, tests can easily be extended by modifying the grammar, without requiring any modifications of the search operators.

\begin{figure}[tb]
  \centering
%  \begin{minipage}[b]{0.3\textwidth}
	\subfloat[Stage with sprites]{\includegraphics[width=0.32\textwidth]{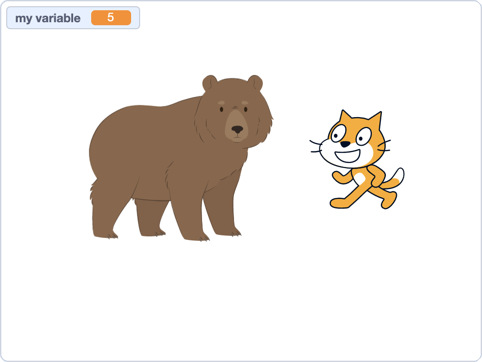}}
	\hfill
	\subfloat[Script of the bear]{\includegraphics[width=0.3\textwidth]{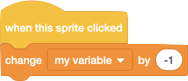}}
%  \end{minipage}
  \hfill
\subfloat[Scripts of the cat]{\includegraphics[width=0.3\textwidth]{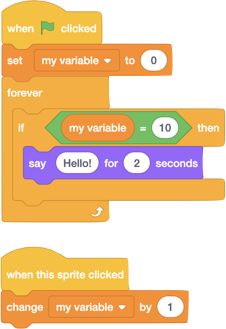}}
    \caption{\label{fig:example1}Example \scratch program: Two sprites controlling the value of a shared variable with \emph{click}-event handlers.}
\end{figure}

As an example of this encoding, consider the \scratch project shown in \cref{fig:example1}, which contains two sprites, a \textit{cat} and a \textit{bear}, and a variable \texttt{my\_variable}.
The cat's script with the green flag hat-block can be seen as the main function of the program, as \scratch programs are started by clicking a green flag in the user interface. When the green flag is clicked, the script initialises the variable \texttt{my\_variable} with value $0$, and then runs a forever-loop. In the forever-loop there is an if-condition which checks if the value of variable \texttt{my\_variable} is $10$; if so, the cat will say ``Hello'' for two seconds. 
Both, bear and cat, have event handlers that are triggered if the user performs a mouse click on either of the sprites. When the user clicks on the \emph{bear} then \texttt{my\_variable} is decremented by $1$, if the user clicks on the \emph{cat} then \texttt{my\_variable} is incremented by $1$.

The available events in this example program are independent of the concrete program state, as there are always only the two event handlers for clicking the two sprites. Thus, assuming a test of length 10, the following grammar describes the possible tests for this program:
{
  $$
      \begin{aligned}
         testcase ::= \; & input_1 \; input_2 \; \ldots \; input_{10} \\
		 input    ::= \; & \textsf{ClickSprite} \; \text{bear} \; | \; \textsf{ClickSprite} \; \text{cat}  \; |  \; \textsf{Wait} \; \text{default}   \; |  \; \textsf{Wait} \; \text{\SI{2}{\second}}
      \end{aligned}
  $$
}%
Here, the $\textsf{Wait} \; \text{\SI{2}{\second}}$ event is based on the \SI{2}{\second} delay in the \emph{say}-block. Consider the following example chromosome in integer encoding:
{
  $$
      \begin{aligned}
         T  = \; & \langle 4 \; 3 \; 5 \; 2 \; 2 \; 1\; 4 \; 6 \; 3 \; 8 \rangle
      \end{aligned}
  $$
}
The decoding would start with symbol $testcase$ and codon $4$. Since there are only $4$ productions for $input$, the decoding to a test case looks as follows:
{
  $$
\begin{array}{lcl}	
4 \; mod \; 4 = 0 & \;\longrightarrow\; & \textsf{ClickSprite} \; \text{bear} \\
3 \; mod \; 4 = 3 & \;\longrightarrow\; & \textsf{Wait} \; \text{\SI{2}{\second}} \\
5 \; mod \; 4 = 1 & \;\longrightarrow\; & \textsf{ClickSprite} \; \text{cat} \\
            & ... & \\
8 \; mod \; 4 = 0 & \;\longrightarrow\; & \textsf{ClickSprite} \; \text{bear}\\
\end{array}
  $$
}

%------------------------------------------------------------------------------
\subsection{Search Operators and Algorithm}
\label{sec:algorithm}

Since the optimal number of events in a test case is problem specific and cannot be known ahead of time, we use a variable length encoding. This also provides an opportunity for the search to minimise the length of the tests.
%, such that users receive shorter tests with less redundancy.
%
To generate a random individual for the initial population we select a random length $n$ in the range $[1..max]$, where $max$ is a predetermined parameter representing the maximum number of events in a test case. Then, we generate $n$ random codons, each of which is selected from the range from $0$ to $480$;
the value $480$ is the width of the stage in pixels and thus the largest possible parameter any of our supported events can take.
The mutation operator can probabilistically (1) replace codons with random codons, (2) insert new codons, and (3) delete codons, each with a certain probability dependent on the length. 
The crossover operator splits the two parent chromosomes into two parts at a randomly selected relative point (i.e., $[0..1]$). Then the codons on the right side of that splitting point are swapped between the chromosomes. 
The mutation and crossover operators are based on prior work on variable size search~\cite{fraser2012whole} with the aim of avoiding test length bloat~\cite{fraser2013handling}.

%\paragraph{Many-objective Search.}

The goal of the optimisation is to produce a set of test cases such that each program statement is covered, thus there is one objective function for each program statement. We therefore use the many-objective sorting algorithm (MOSA)~\cite{MOSA}, which overcomes the scalability problems of traditional many-objective algorithms. 
The initial random population evolves toward better populations through subsequent generations until a stopping condition is reached. In each generation an offspring population of the same size is created by selecting test cases from the parent population and modifying them using crossover and mutation. For this, rank selection is used, which gives better test cases a higher probability of being selected. 

During evolution, the test cases in the parent and offspring populations are classified into different fronts. At first, for each uncovered statement, the shortest test case, that is closest to covering the statement, is computed and added to $front_0$. The remaining test cases are sorted according to Pareto dominance: A test case $x$ dominates another test case $y$, if $x$ is better or equal than $y$ for all uncovered statements and better than $y$ for at least one uncovered statement. All test cases not dominating each other are assigned the same front. The obtained Pareto fronts \emph{front}$_1$, $\ldots$, \emph{front}$_n$ are sorted in descending order by dominance, meaning test cases in lower fronts dominate test cases in higher fronts. After ranking all test cases a new parent population is formed by adding the test cases in \emph{front}$_0$ gradually followed by the subsequent fronts until the population size is reached. If the front is too big to add all test cases, a test case with a greater distance to other test cases is preferred in order to promote diversity. With this approach the search focuses towards the uncovered statements, because the best test cases for these statements are likely to survive. Furthermore, there exists an archive, which stores the shortest covering test case for each covered statement. The test cases in the archive are updated in each generation and form the test suite at the end of the algorithm.
%------------------------------------------------------------------------------
\subsection{Fitness Function}
\label{sec:fitness}

As basic coverage criterion we consider statement coverage, such that for each statement in a program under test we derive a separate fitness function. The fitness function for a given target statement is encoded using the traditional combination of approach level~\cite{WBS01} and branch distance~\cite{Kor90}. Given a test $t$ and target statement $s$, the fitness function is defined as:
{
\begin{equation*}
  f(t, s) = approachLevel(t, s) + \alpha (branchDistance(t, s))
\end{equation*}
}%
 where $\alpha$ denotes a normalisation function in the range $[0..1]$~\cite{arcuri_it_2013}.

\begin{figure}[tb]
\includegraphics[width=\textwidth]{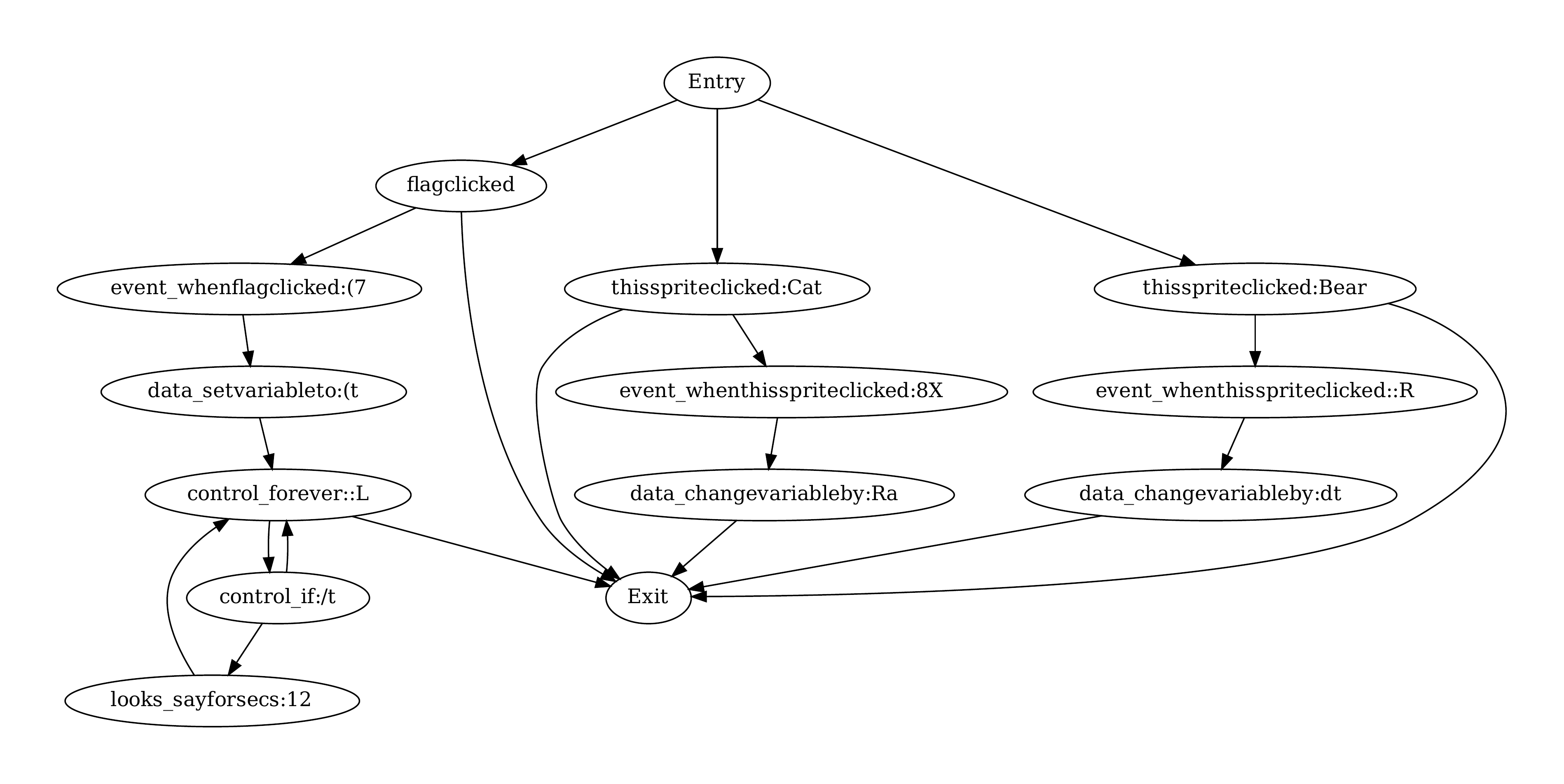}
\caption{Control Flow Graph created for the example program from \cref{fig:example1}.}\label{fig:cfg}
\end{figure}

% \begin{figure}[tb]
% \includegraphics[width=\textwidth]{figures/cdg}
% \caption{Control Dependence Graph created for the CFG in \cref{fig:cfg}.}\label{fig:cdg}
% \end{figure}

% A \scratch program consists of the \emph{stage}, which represents the application window and background image, and a collection of \emph{sprites} that are rendered as different images on top of the stage. The stage and each of the sprites contain a number of \emph{scripts} $\scripts$ that define the program logic. Scripts are created by visually arranging \emph{blocks} that correspond to syntactical elements of the language, such as control-flow structures or expressions.
%

A given target \scratch program consists of a number of scripts $\scripts$ (see \cref{sec:background}); for each script $s = (\ctrllocations, \datalocations, \controlflows, l_0) \in \scripts$ we derive the control flow graph (CFG), defined as $CFG = (\ctrllocations \cup \{entry, exit\}, \controlflows)$, i.e., a directed graph consisting of nodes $\ctrllocations$ as well as dedicated $entry$ and $exit$ nodes,  and edges $\controlflows$. We combine these intraprocedural CFGs to an interprocedural super-CFG as follows:

\begin{itemize}
	\item For each event handler, we add an artificial node with edges to the event handler (\emph{hat block}) as well as the $exit$ node, effectively turning event nodes into branching nodes in the CFG. We further add an edge from $entry$ to this artificial node for event handlers of user inputs. 
	\item For each \emph{broadcast} statement, we add an edge from the broadcast to all scripts that start with a matching receive event handler block.
	\item For each \emph{create clone} statement, we add an edge from the \emph{create clone} block to all scripts that start with a matching \emph{When I start as clone} event handler block for the corresponding sprite.
	\item For each \emph{procedure call} statement, we add an edge from the call to the start block of the procedure (custom block), and a return edge from the end of the procedure to the successor node in the calling script.
\end{itemize}

Figure~\ref{fig:cfg} shows the interprocedural CFG for the program in \cref{fig:example1}. This CFG contains three artificial event nodes (\textit{thisspriteclicked:Cat}, \textit{thisspriteclicked:Bear} or \textit{flagclicked}), each of which effectively is a branching statement depending on whether the event occurs. These branches turn the occurrence of events into control dependencies of the statements in the event handler code.
%, as shown in the control dependence graph (CDG, \cref{fig:cdg}).

%When a fitness function is initialised for a target statement we pre-calculate the approach levels for all its control  dependency nodes.This is done by walking backwards from the target node to the start node and storing all control dependent nodes in a map with the respective approach levels. 
In order to measure the fitness, we instrument program executions to produce traces of the branching statements executed. Given a trace, the control dependence graph is used to calculate the approach level for a given target node. 
%For example, for the node \textit{looks\_sayforsecs:12} in Figure~\ref{fig:cdg}, if the execution reached the parent node \textit{control\_if:\/t} then the approach level is $0$, and if only \textit{contorl\_forevel::L} was reached then the approach level is $1$.
%
For each branch, the execution trace further contains information about the minimum branch distances (for evaluation to true and to false). For the fitness evaluation we then use the minimum branch distance of the branching node with the lowest approach level for our target node.

A particularly interesting aspect of \scratch programs is that predicates in the code often refer to the locations and interactions of the sprites on the stage, in particular to check whether a sprite touches another sprite.
We instrument the corresponding \emph{reporter block} such that it produces not just a binary true/false result, but an actual distance measurement. 
In case the sprites are touching the branch distance for the true evaluation is $0$ by definition; if they are not touching we use the \textit{distanceTo} function in \scratch to determine the distance between the sprites, and use that as the branch distance.
Similarly, if the condition checks if a sprite is touching the edge of the stage, we can gather the position information and calculate the distance to all edges, and use the minimal distance as the branching distance. Further predicates (e.g., \emph{touching colour}) can be approximated with branch distances similarly.

%------------------------------------------------------------------------------
\subsection{Headless Accelerated Test Execution}
\label{sec:execution}

\scratch tests are executed by running a \scratch program and applying events encoded in a test to the \scratch VM.
This kind of test mimics a normal execution of \scratch, which is as fast as the regular execution when a user runs the program.
The result is that running tests can be very time consuming.
We added two modifications to the \scratch VM to increase its execution speed, which in turn decreases the time to execute tests.

% Removed as this is a double blind study
%Niklas Zantner realized such an acceleration as part of his Master Thesis \cite{Zantner20}.

The \scratch VM updates its internal state and the UI representation with a given interval.
The first modification introduces an acceleration factor which reduces the default update interval.
By simply reducing the interval, the VM updates its state more often which leads to faster execution of regular blocks.

However, only increasing the number of state updates of the VM is not sufficient:
Blocks that use time (e.g., waiting for $x$ seconds) measure time not in state updates but in real time, and simply accelerating the state updates does not speed up the perceived time of blocks.
A block that is waiting for 2 seconds will still wait 2 seconds even if we speed up that execution by a factor of 10.

To solve this problem we instrumented blocks that use time to speed up their waiting time according to the acceleration factor.
The following example shows the waiting function of the \textit{control\_wait} \scratch block.
By default, the control block makes the application wait for the amount of seconds defined in \texttt{args.DURATION}:

\begin{lstlisting}[basicstyle=\small\ttfamily,language=ES6,tabsize=2]
wait (args, util) {
  if (util.stackTimerNeedsInit()) {
      const duration = Math.max(
        0, 
        1000 * Cast.toNumber(args.DURATION)
      );
  ...
\end{lstlisting}

To reduce the waiting time of the block the \texttt{args.DURATION} has to be reduced by the acceleration factor. 
For example, to run the application with an acceleration factor of 10 instead of 1, the duration has to be divided by factor 10.
Instrumentation of this function is achieved by dynamically replacing the function with a modified version as follows:

\begin{lstlisting}[basicstyle=\small\ttfamily,language=ES6,tabsize=2]
const accelerationFactor = 10;
const original = this.vm.runtime._primitives.control_wait;
const instrumented = (args, util) => {
    const clone = {...args};
    clone.DURATION = args.DURATION / accelerationFactor;
    return original(clone, util);
};
this.vm.runtime._primitives.control_wait = instrumented;
\end{lstlisting}

%With all necessary modifications in place, the test execution could be accelerated by a factor of 10.
%\todo{describe what could cause problems and why we cannot accelerate by 50 or 100?}

%This update frequency-based approach was one of the two results used for the search-based testing of the thesis.
%The second result will be described in the following section.

As a third modification the \servant front-end was created for \whisker.
The \servant is a Node.js based command-line interface (CLI) which is based  on \emph{Puppeteer}\footnote{\url{https://pptr.dev/}, last accessed 9.6.2020}, a \textit{Headless Chrome Node.js API}.
%The servant wraps \whisker to manage Scratch application testing.
This new front-end makes it possible to use \whisker in headless environments (i.e., without a graphical user interface), so that tests can be executed on computing clusters.

%To allow Whisker users to execute Whisker within headless environments and via a more abstract interface, the Servant was created.
%Technically, the Servant is  which is provided by the Chrome DevTools team as an open-source project via GitHub.
%Puppeteer allows Whiskert users to execute and control a Google Chrome instance, including the \texttt{--headless} flag of Google Chrome.
%The headless flag enables developers to run Google Chrome in headless environments, allowing users to execute HTML, CSS, and JavaScript code in a browser without rendering.
%
%The following example showcases how the Servant can be used to trigger a search-based test from the command line:
%\begin{lstlisting}[basicstyle=\scriptsize, language=BASH]
%node servant -s ./CatchGame.sb3 -c ./mosa.json -a 10 -d -g
%\end{lstlisting}
%The command will execute the search-based test defined in the \texttt{./mosa.json} config on the project defined in \texttt{./CatchGame.sb3}, with an acceleration factor of 10 (\texttt{-a 10}), in the headless mode (\texttt{-d} as in \textit{decapitated}).

%%%%%%%%%%%%%%%%%%%%%%%%%%%%%%%%%%%%%%%%%%%%%%%%%%%%%%%%%%%%%%%%%%%%%%%%%%%%%%%
\section{Case Study Examples}
\label{sec:evaluation}

\begin{figure}[t]
	\subfloat[\label{fig:pingpong_stage}Ping Pong]{\includegraphics[width=0.3\textwidth]{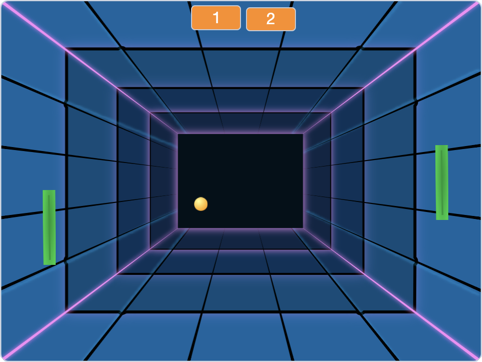}}
	\hfill
	\subfloat[\label{fig:fruit_stage}Fruit Catching]{\includegraphics[width=0.3\textwidth]{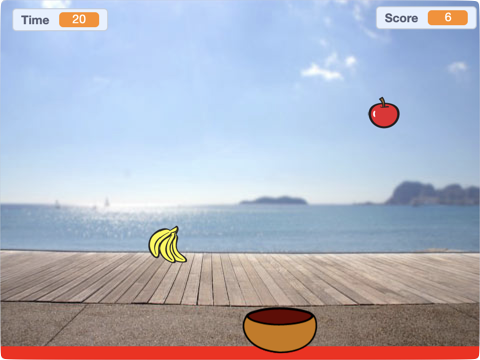}}
	\hfill
	\subfloat[\label{green_stage}Green Your City]{\includegraphics[width=0.3\textwidth]{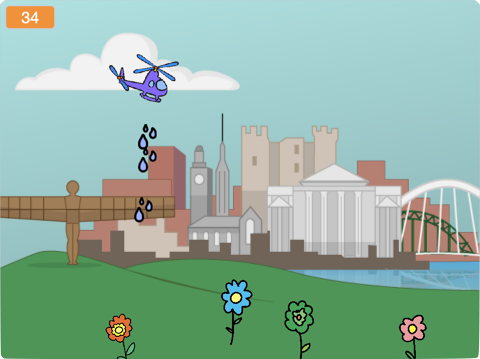}}
	\\
	\subfloat[\label{fig:fruit_code}Code excerpt of the Fruit Catching game]{\includegraphics[trim=0 65 0 275,clip,width=0.35\textwidth]{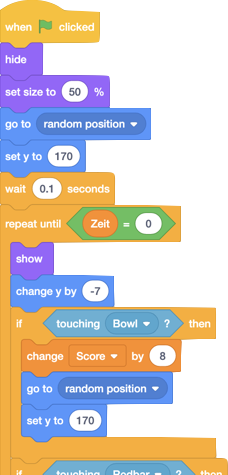}}	
	\subfloat[\label{fig:green_code}Code excerpt of the Green Your City game]{\includegraphics[trim=0 80 0 50,clip,width=0.6\textwidth]{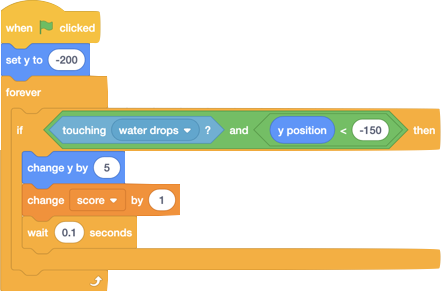}}	
%	\subfloat[\label{fig:fruit_code}Code excerpt of the Fruit Catching game]{\includegraphics[trim=0 200 0 140,clip,width=0.32\textwidth]{figures/fruit_code}}	
	\\
	\subfloat[\label{fig:pingpong_code}Code excerpt of the Ping Pong game]{\includegraphics[trim=0 300 0 140,clip,width=\textwidth]{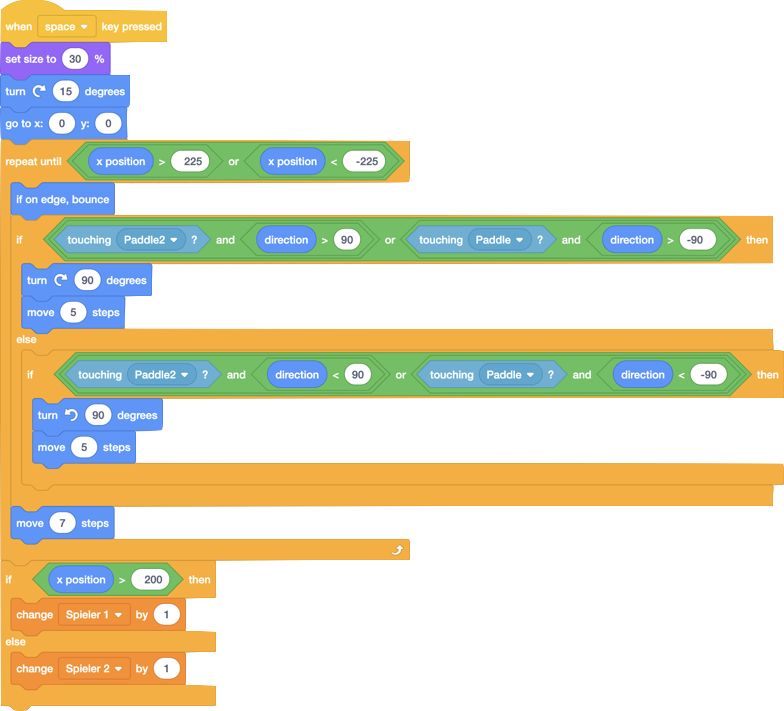}}
	
	\caption{\label{sec:examples}Case study programs and code examples.}
\end{figure}

To illustrate the proposed approach, we use three example \scratch programs
shown in \cref{sec:examples}. The Ping Pong game is a classical \scratch
tutorial example, and represents the types of programs commonly created by
early learners. It consists of 36 blocks in 4 scripts, and is controlled by
five different key events. We modified the Ping Pong game to initialise all
variables (missing variable initialisation is a common bug in
\scratch~\cite{fraedrich2020}). The Fruit Catching game was used by Stahlbauer
et al.~\cite{TestingScratch} as part of their evaluation of \whisker, and is
taken from an educational context. It consists of 49 blocks in 4 scripts, and
is controlled by the cursor keys. Green Your City is part of the popular CodeClub collection of example projects. It consists of 8 scripts and 52 blocks.

\FloatBarrier

\begin{figure}[t]
	\subfloat[\label{fig:banana1}Branch distance 284]{\includegraphics[width=0.31\textwidth]{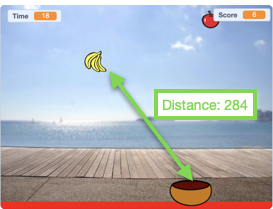}}
	\hfill
	\subfloat[\label{fig:banana2}Branch distance 188]{\includegraphics[width=0.31\textwidth]{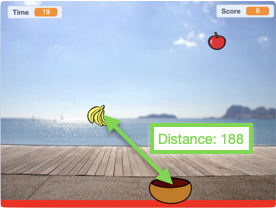}}
	\hfill
	\subfloat[\label{fig:banana3}Branch distance 87]{\includegraphics[width=0.31\textwidth]{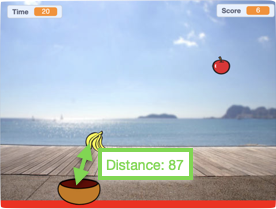}}	
	\caption{\label{fig:banana_distance}Fitness values for statements within the if-condition shown in \cref{fig:fruit_code}.  \\
	{\small The approach level is 0 in all cases, but the branch distance of the if-condition depends on the distance between the bowl and the banana.}}
\end{figure}

\begin{figure}[t]
	\subfloat[\label{fig:pingpong1}Branch distance 190]{\includegraphics[width=0.31\textwidth]{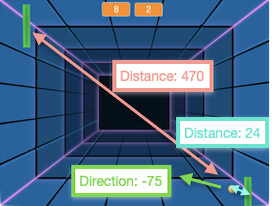}}
	\hfill
	\subfloat[\label{fig:pingpong2}Branch distance 224]{\includegraphics[width=0.31\textwidth]{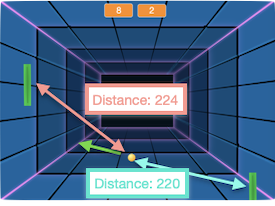}}
	\hfill
	\subfloat[\label{fig:pingpong3}Branch distance 37]{\includegraphics[width=0.31\textwidth]{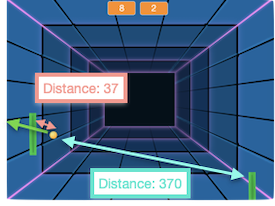}}	
	\caption{\label{fig:pingpong_distance}Fitness values for statements within the first if-condition shown in \cref{fig:pingpong_code}.  \\
	{\small The approach level is 1 if $x > 225$ or $x < -225$, otherwise it is 0. Given that the ball faces in direction $-75$ (such that the branch distance for $-75 > 90$ is $90 - -75 + 1 = 166$, and the branch distance for $-75 > -90$ is $0$) and is between the paddles, the branch distance is calculated as $min(distance(ball, right\_paddle) + 166, distance(ball, left\_paddle) + 0))$.}}
	\vspace{-1em}
\end{figure}
 
\begin{figure}[t]
	\subfloat[\label{fig:green1}Branch distance 136]{\includegraphics[width=0.31\textwidth]{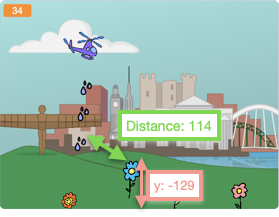}}
	\hfill
	\subfloat[\label{fig:green2}Branch distance 128]{\includegraphics[width=0.31\textwidth]{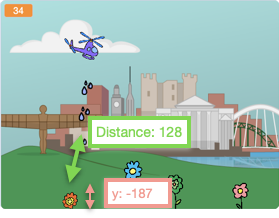}}
	\hfill
	\subfloat[\label{fig:green3}Branch distance 297]{\includegraphics[width=0.31\textwidth]{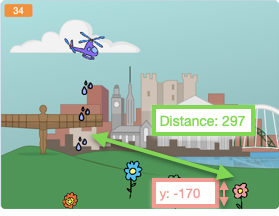}}	
	\caption{\label{fig:green_distance}Fitness values for statements within the first if-condition shown in \cref{fig:green_code}, which is duplicated for each flower. \\
	{\small The approach level is 0 if the green flag is clicked; the branch distance is calculated based on the $y$-position and the distance to the closest water drop. Given that a flower needs to be below $y = -150$, the branch distance for the flower in \cref{fig:green1} is $distance(flower, water) + (-129 - -150 + 1) =  136$. The flower in \cref{fig:green2} is below $y = -150$, therefore the branch distance is $distance(flower, water) + 0 = 128$, and for \cref{fig:green3} it is 297.}}
	\vspace{-1em}
\end{figure}

% \begin{figure}[t]
% 	\centering
% 	\includegraphics[width=0.8\textwidth]{figures/fruit_evolution}
% 	\caption{\label{fig:fruit_evolution}Example evolution (1+1 EA) for the fitness function of catching the banana (Code in \cref{fig:fruit_code}, fitness examples in \cref{fig:banana_distance}) and corresponding chromosomes. After the banana has been caught at iteration 15, the evolution continues shortening the test case down to pressing the right key three times, and then left two times.}
% 	\vspace{-1em}
% \end{figure}

\FloatBarrier

\begin{figure}[!t]
	\subfloat[\label{fig:result_pingpong}Ping Pong]{\includegraphics[width=0.33\textwidth]{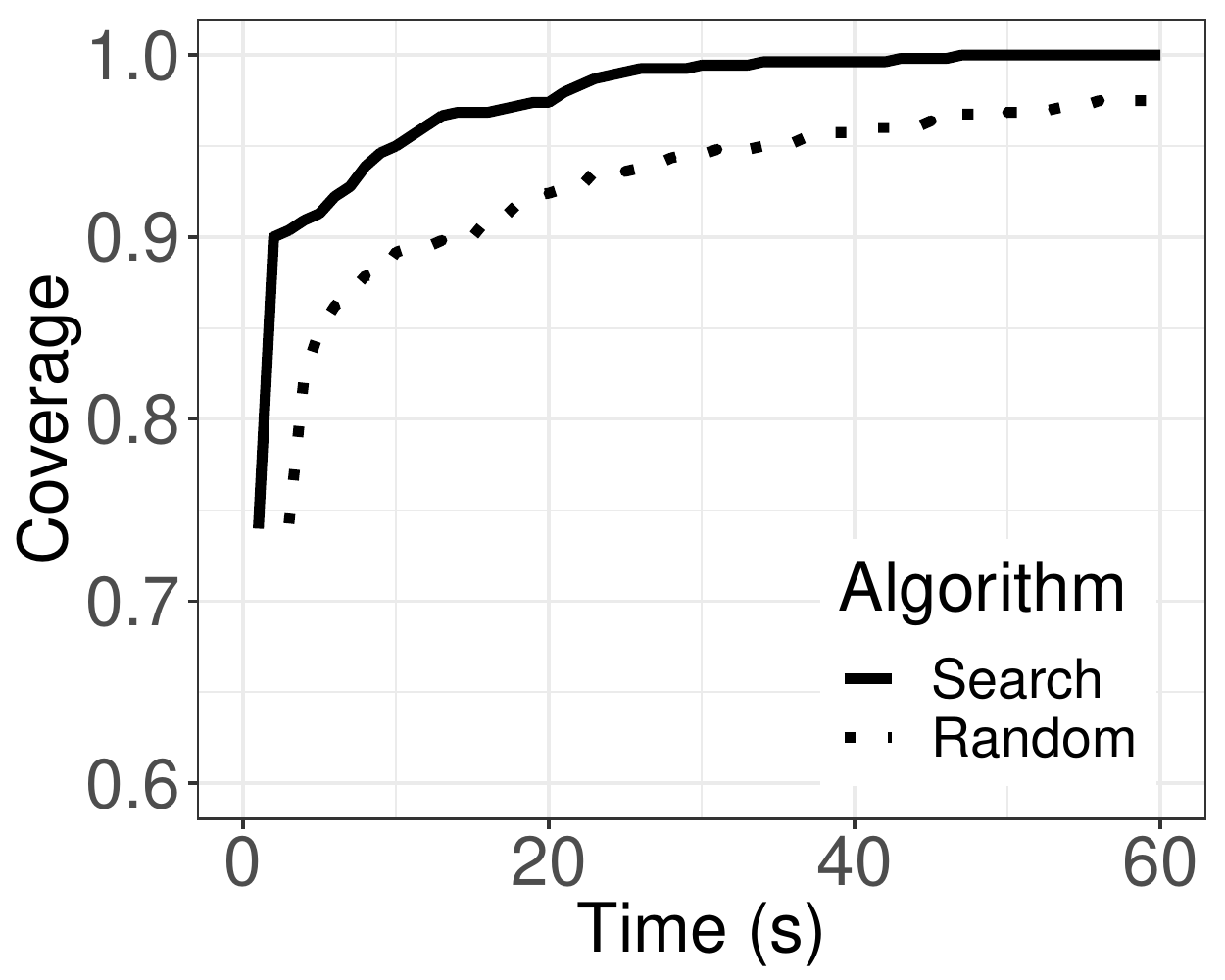}}
	\hfill
	\subfloat[\label{fig:result_fruit}Fruit Catching]{\includegraphics[width=0.33\textwidth]{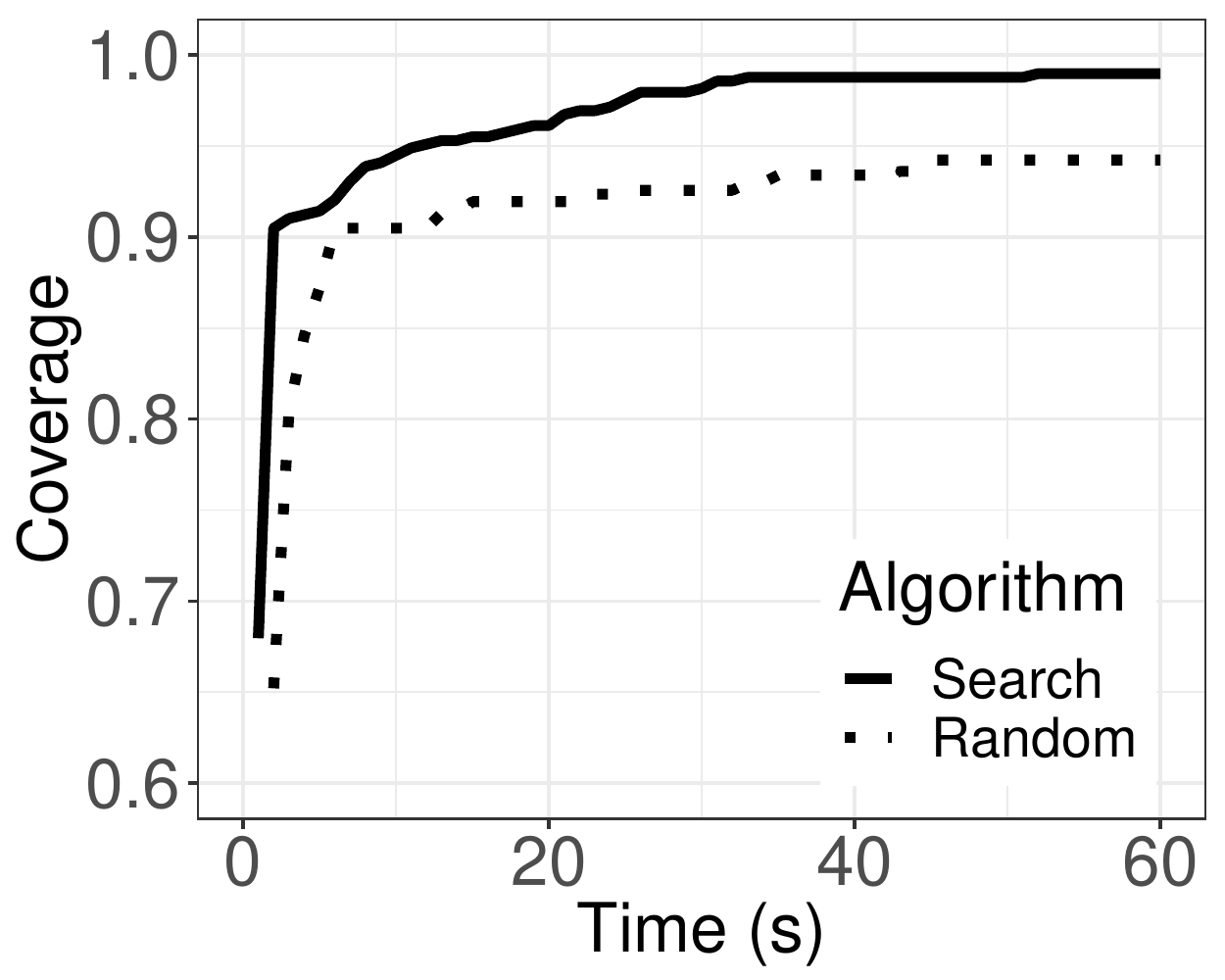}}
	\hfill
	\subfloat[\label{fig:result_green}Green Your City]{\includegraphics[width=0.33\textwidth]{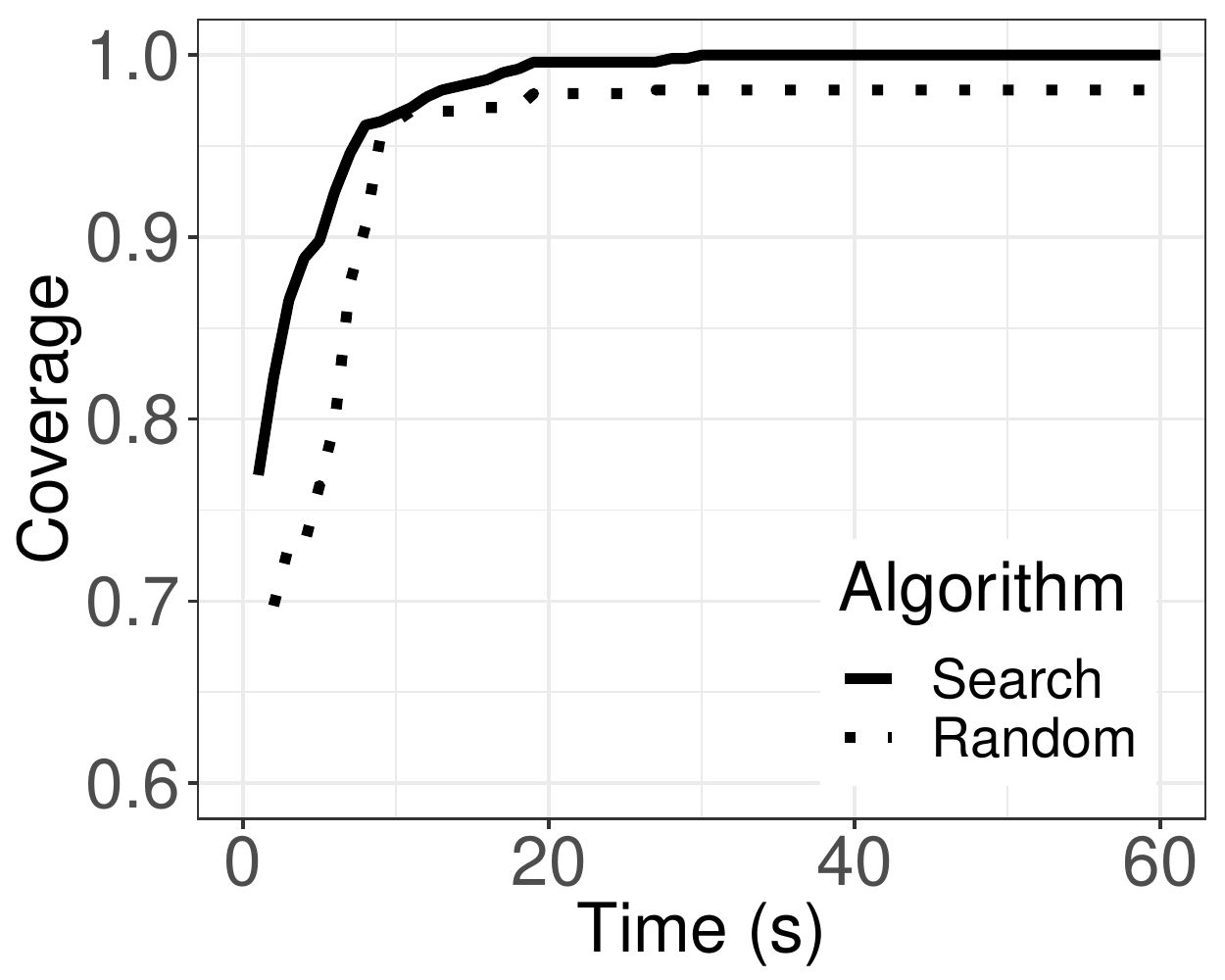}}
	\caption{\label{fig:coverage}Coverage over time, averaged over 30 runs.}
\end{figure}

In a typical \scratch program, many statements are trivially covered, but there can always be tricky code that poses challenges for automated testing.  
\Cref{fig:fruit_code} shows an excerpt of the code of the Fruit Catching game, taken from the banana sprite: In order to execute the statements in the then-block of the shown branch, the player-controlled bowl needs to touch the banana. The \emph{touching} predicate is reflected by a branch distance that guides the test generator towards achieving this, as illustrated in \cref{fig:banana_distance} for different states of the game. %\Cref{fig:fruit_evolution} demonstrates the evolution for one coverage goal in the then-branch in terms of the fitness values and corresponding chromosomes.

The if-condition of the Ping Pong game shown in \cref{fig:pingpong_code} requires a branch distance calculation by applying the rules on nested logical expressions to combine the distances between the ball and the two paddles, as well as the required and current orientation of the ball. As suggested by the example states shown in \cref{fig:pingpong_distance}, the fitness landscape induced by the complex expression is less convenient than in the simple \emph{touching}-predicate shown in \cref{fig:banana_distance}. Green Your City also contains logical expressions combining various distance measurements; in particular each of the flowers contains the code shown in \cref{fig:green_code}. Flowers grow vertically each time they touch a water drop, the example branch is covered once this has happened sufficiently often to let the flower grow above $y = -150$ and it touches a water drop again; \cref{fig:green_distance} shows examples of this distance for different flowers.

To see the search in action, \Cref{fig:coverage} shows the coverage over time,
averaged over 30 runs of \whisker using the original
configuration~\cite{TestingScratch} (``Random'', i.e., \SI{60}{\second}, events every \SI{250}{\milli\second},
program reset every \SI{10}{\second}), and the search extension (``Search'', \SI{60}{\second} runtime,
population size 10, \SI{250}{\milli\second} default event duration, acceleration factor 5, crossover probability 0.8, initial length 2).
The improvement of the search over random testing after \SI{60}{\second} is statistically significant with $p < 0.001$ in all three cases: For Ping Pong the Vargha-Delaney effect size $\hat{A}_{12} = 0.7$, for the Fruit Catching game $\hat{A}_{12} = 1.0$, and for Green Your City $\hat{A}_{12} = 1.0$.
For Ping Pong (\cref{fig:result_pingpong}) the search covers most statements quickly, but generally needs longer ($\approx$\SI{30}{\second}) to cover the two branches discussed in example \cref{fig:pingpong_distance}. Random testing sometimes hits the ball by accident, but generally needs to play much longer before that happens. Similarly, for the Fruit Catching game \cref{fig:result_fruit} shows, while again many statements are easy to cover, the search needs about \SI{30}{\second} to succeed in catching both types of fruit, while random takes substantially longer.
The results for Green Your City (\cref{fig:result_green}) are similar, in that the search requires around \SI{30}{\second} until all statements are covered, while random usually takes longer; the tricky branches are those requiring to hit flowers with water drops, as well as one statement that requires the helicopter to reach the bottom, and thus a longer play time than random testing used would be necessary. While these are promising initial results, a more thorough evaluation on a larger number of \scratch projects is planned as future work. The randomised and game-like nature of most \scratch projects can be a cause for test flakiness; however, this flakiness can be easily contained~\cite{TestingScratch}.

Note that an automatically generated coverage test suite is only one of the stepping stones towards providing automated feedback to learners: The generated tests tend to be short and their randomised nature potentially makes them difficult to understand, and so we do not expect that the tests would be shown \emph{directly} to learners. The tests are rather meant as input to further dynamic analysis tools, for example in order to serve as test harness for \whisker~\cite{TestingScratch}, where manually written \scratch observers would provide insight into which aspects of the functionality are correctly implemented in a concrete learner's implementation of a predefined programming challenge. We envision that the concrete feedback to learners will be provided in terms of textual hints or code suggestions~\cite{price2017isnap,techapalokul2019code}, and the generated tests are a prerequisite for achieving this.

%%%%%%%%%%%%%%%%%%%%%%%%%%%%%%%%%%%%%%%%%%%%%%%%%%%%%%%%%%%%%%%%%%%%%%%%%%%%%%%
\section{Related Work}
\label{sec:related}

The rising popularity of block-based programming languages like \scratch creates an increased demand for automated analysis to support learners. It has been shown that learners tend to adopt bad habits~\cite{meerbaum2011habits} and create ``smelly'' code~\cite{aivaloglou2016kids,hermans2016smells,techapalokul2017understanding}; these code smells have been shown to have a negative impact on understanding~\cite{hermans2016code}. To counter this, automated tools can help to identify and overcome such problems.
For example, the \textsc{Dr. Scratch}~\cite{moreno2015dr} website points out a small number of code smells to learners, using the \textsc{Hairball}~\cite{boe2013hairball} static analysis tool, and similar smells are identified by \textsc{Quality hound}~\cite{techapalokul2017quality} and \textsc{SAT}~\cite{chang2018scratch}. 
\textsc{LitterBox}~\cite{fraedrich2020} can identify patterns of common bugs in \scratch programs. 
Besides pointing out possible mistakes, it is desirable to also identify helpful suggestions and feedback, such as what step to take next~\cite{price2017isnap} or how to remove code smells~\cite{techapalokul2019code}.
The majority of existing approaches are based on static program analysis, and can therefore only provide limited reasoning about the actual program behaviour. The \textsc{Itch} tool~\cite{johnson2016itch} translates a 
small subset of \scratch programs to Python programs (textual interactions 
via say/ask blocks) and then allows users to run tests on these programs.  \whisker~\cite{TestingScratch} takes this approach a step further and, besides execution of automated tests  directly in \scratch, also provides automated property-based testing. 

We introduce search-based testing as a means to fully automate the generation of test suites for \scratch programs. These test suites are intended to be the input to dynamic analysis tools that can then use the dynamic information to produce hints and feedback. 
Our approach is based on evolutionary search, which is common for API-level test generation~\cite{fraser2012whole}, but has also been applied to GUI testing~\cite{mao2016sapienz,mahmood2014evodroid,gross2012search}. The concept of grammatical evolution~\cite{o2001grammatical} has not been thoroughly explored in the context of test generation yet~\cite{anjum2020seeding}.

%%%%%%%%%%%%%%%%%%%%%%%%%%%%%%%%%%%%%%%%%%%%%%%%%%%%%%%%%%%%%%%%%%%%%%%%%%%%%%%
\section{Conclusions}
\label{sec:conclusions}

In this paper, we have introduced the idea to apply search-based testing for the problem of generating coverage-oriented test suites for \scratch programs. The use of Grammatical Evolution allows a clean separation between aspects of the meta-heuristic search, and the technical challenges posed by the testing environment. The specific graphical nature of \scratch programs provides opportunities for guidance beyond those common in regular programs. Our extension of the \whisker test generator has demonstrated its potential on a number of example programs. However, there are remaining challenges to be addressed in future work, such as refined support for all user actions, better integration of seeding, consideration of program state (which, for example, is often encoded by the costumes/backdrops in use), and many others, permitting a larger scale evaluation. Furthermore, future work will be able to build on the test suites generated by our approach for further analysis and for generating actionable feedback to users. For example, we anticipate that an example application scenario will be that where a teacher produces a golden solution, then generates a test suite for it, and this test suite then serves for fault localisation or repair suggestions. To support this future work, our extensions to \whisker are available as open source at:
\begin{center}
	 \texttt\url{https://github.com/se2p/whisker-main}
\end{center}

\section*{Acknowledgements}
This work is supported by EPSRC project EP/N023978/2 and DFG project
FR 2955/3-1  ``TENDER-BLOCK: Testing, Debugging, and Repairing
Blocks-based Programs''.

%
% ---- Bibliography ----
%
\bibliographystyle{splncs04}
\bibliography{related}
\end{document}